\documentclass[traditabstract]{aa}
\usepackage{graphicx,epsf,textcomp,pstricks,psfrag,color}
\usepackage{txfonts}
\usepackage{longtable,booktabs}
\usepackage{natbib}
\bibpunct{(}{)}{;}{a}{}{,}

\begin{document}

\sloppy

%-----------------------------------------------------------------------
% Authors' Macros
%-----------------------------------------------------------------------

\newcommand{\be}{\begin{equation}}
\newcommand{\ee}{\end{equation}}
\newcommand{\bd}{\begin{displaymath}}
\newcommand{\ed}{\end{displaymath}}
\newcommand{\bea}{\begin{eqnarray}}
\newcommand{\eea}{\end{eqnarray}}
\newcommand{\etal}{et al.}

\newcommand{\mum}{\,\mu\hbox{m}}
\newcommand{\mm}{\,\hbox{mm}}
\newcommand{\cm}{\,\hbox{cm}}
\newcommand{\m}{\,\hbox{m}}
\newcommand{\km}{\,\hbox{km}}
\newcommand{\AU}{\,\hbox{AU}}
\newcommand{\second}{\,\hbox{s}}
\newcommand{\yr}{\,\hbox{yr}}
\newcommand{\g}{\,\hbox{g}}
\newcommand{\kg}{\,\hbox{kg}}
\newcommand{\rad}{\,\hbox{rad}}
\newcommand{\erg}{\,\hbox{erg}}

\newcommand{\vecbold}[1]{{\bf #1}}

%-----------------------------------------------------------------------
% Title
%-----------------------------------------------------------------------

\title{Can gas in young debris disks be constrained\\
by their radial brightness profiles?
}

\author{Alexander V. Krivov\inst{1}
        \and
        Fabian Herrmann\inst{1}
        \and
        Alexis Brandeker\inst{2}
        \and
        Philippe Th{\' e}bault\inst{3,2}
       }
\offprints{A.V.~Krivov, \email{krivov@astro.uni-jena.de}}
\institute{Astrophysikalisches Institut und Universit\"atssternwarte, Friedrich-Schiller-Universit\"at 
Jena,\\
           Schillerg\"a{\ss}chen~ 2--3, 07745 Jena, Germany
          \and
          Department of Astronomy, Stockholm University, SE-106\,91 Stockholm, Sweden
          \and
          LESIA, Observatoire de Paris,
          F-92195 Meudon Principal Cedex, France
          }
\date{Received July 17, 2009; accepted {September 15, 2009}}

\abstract
{
Disks around young stars are known to evolve from optically thick,
gas-dominated protoplanetary disks to optically thin, almost gas-free debris disks.
It is thought that the primordial gas is largely removed
at ages of $\sim 10$~Myr and indeed, only little amounts of gas
have been deduced from observations for debris disks at ages of $\ga 10$~Myr.
However, gas detections are  difficult and often indirect,
not allowing one to discern the true gas densities.
This suggests using dynamical
arguments: it has been argued that gas, if present with higher densities,
would lead to flatter radial profiles of the dust density and brightness
than those actually observed. In this paper, we systematically study
the influence of gas on the radial profiles of brightness.
We assume that dust is replenished by planetesimals orbiting in a
``birth ring'' and model the dust distribution and scattered-light brightness profile
in the outer part of the disk exterior to the birth ring, under different
assumptions about the gas component.
Our numerical simulations, supported with an analytic model,
show that the radial profile of dust density and the surface brightness
are surprisingly insensitive to variation of the parameters of a central star,
location of the dust-producing planetesimal belt,
dustiness of the disk and~--- most importantly~--- the parameters of the ambient gas.
The radial brightness slopes in the outer disks are all typically
in the range $-3$...$-4$.
This result holds for a wide range of gas densities (three orders of magnitude),
for different radial profiles of the gas temperature,
both for gas of solar composition and gas of strongly non-solar composition.
The slopes of $-3$...$-4$ we find are the same
that were theoretically found for gas-free debris disks,
and they are the same as actually retrieved from observations
of many debris disks.
Our specific results for three young (10--30~Myr old), spatially resolved,
edge-on debris disks  ($\beta$~Pic, HD~32297, and AU Mic) show
that the observed radial profiles of the surface brightness do not pose
any stringent constraints on the gas component of the disk.
We cannot exclude that outer parts of the systems
may have retained substantial amounts of primordial gas
which is not evident in the gas observations
(e.g. as much as $50$ Earth masses for $\beta$~Pic).
However, the possibility that gas, most likely secondary,
is only present in little to moderate
amounts, as deduced from gas detections
(e.g. $\sim 0.05$ Earth masses in the $\beta$~Pic disk or even less),
remains open, too.

\keywords{planetary systems: formation --
          circumstellar matter --
          celestial mechanics --
          stars: individual: $\beta$~Pic --
          stars: individual: HD~32297 --
          stars: individual:  AU Mic.
         }
}
\authorrunning{Krivov et al.}
\titlerunning{Gas in young debris disks}

\maketitle

%-----------------------------------------------------------------------
% Main Text
%-----------------------------------------------------------------------

\section{Introduction}
 
In the course of their evolution, circumstellar disks transform from optically thick,
gas-dominated protoplanetary disks to optically thin, almost gas-free debris disks.
How the gas is removed is not known in detail, but the removal is thought to be the result
of a quick inside-out process at an age of $\sim 10$~Myr
\citep{alexander-2008,hillenbrand-2008}.
It may be either
due to the UV switch mechanism resulting from an interplay between
photoevaporation and viscous accretion
\citep[e.g.][]{hollenbach-et-al-2000,clarke-et-al-2001,%
takeuchi-lin-2005,takeuchi-et-al-2005,alexander-armitage-2007}
or due to gap opening by hidden giant planets
\citep[e.g.][]{lubow-et-al-1999,lubow-dangelo-2006}.
Although the former effect seems to be slightly preferred by observational
statistics \citep[e.g.][]{cieza-et-al-2008},
it is not yet possible to distinguish between them with certainty
\citep{najita-et-al-2007,alexander-2008,hillenbrand-2008}.

In this paper, we concentrate on a more advanced phase of system's
evolution: the debris disk stage. Apart from possible planets, a debris disk
system contains remnant planetesimals and dust into
which they are ground through collisions \citep[see, e.g.][and references therein]{wyatt-2008}.
Debris disks are expected to be nearly gas-free, at least extremely
gas-poor compared to protoplanetary disks. Even so, in the case of
$\beta$~Pic, gas was detected very early on 
in absorption
\citep{slettebak-1975,hobbs-et-al-1985}, and later in emission
\citep{olofsson-et-al-2001}, due to the favorable edge-on orientation
of the disk.
The observed gas around $\beta$~Pic is most likely replenished,
i.e.\ secondary, as opposed to a remnant from the initial star-forming
cloud. Evidence for this comes from the presence of CO
\citep{vidalmadjar-et-al-1994,jolly-et-al-1998,roberge-et-al-2000}, which
would be dissociated on time-scales of $\sim$200\,yr
\citep{vandishoeck-black-1988,roberge-et-al-2000}
and from the presence of neutral gas elements in the disk
 \citep{olofsson-et-al-2001,brandeker-et-al-2004}, subject to short removal
times \citep{fernandez-et-al-2006}.
Possible mechanisms for producing secondary gas include
photon-induced desorption from solids
\citep{chen-et-al-2007} and grain-grain collisions
\citep{czechowski-mann-2007}.
Part of the observed gas may also stem from
comet evaporation, as inferred from observed time-variable absorption lines
\citep[e.g.][]{ferlet-et-al-1987,beust-valiron-2007}.

However, in general observations of gas are much more difficult than that of dust.
Standard detection techniques either use CO as a tracer of hydrogen which can be
observed at radio frequencies, as done by \citet{hughes-et-al-2008} for 49~Cet,
or measure $\mathrm{H}_2$ emission lines which are pumped by stellar
emission lines originating from the chromospheric and coronal regions, which
was done for AU~Mic by \citet{france-et-al-2007}. A potentially more sensitive way of
finding gas is to look for it in absorption, as was done for $\beta$~Pic. The 
downside is that this requires the special edge-on geometry of the disk, but this
method has nevertheless been successfully used by  \citet{redfield-2007} to
detect circumstellar Na\,I absorption towards HD~32297, a star with a known 
disk. Conversely, stars which are known to exhibit circumstellar absorption lines,
so called \textit{shell stars}, can be searched for evidence of circumstellar
material, as done by \citet{roberge-weinberger-2008} using Spitzer/MIPS data.
Out of 16 surveyed shell stars they found infrared excess, and thus evidence for
circumstellar dust, around four stars: HD\,21620, HD\,118232, HD\,142926 and
HD\,158352.

Despite substantial efforts, the gas component of the debris disks
remains much less constrained observationally than the dusty one.
It is quite possible that primordial gas survives longer than
usually assumed, at least in the outer parts of the disks, or is present in larger
amounts than expected, without showing up in observations.
In fact, about ten Earth masses
of gas, if not more, could still remain in many young debris disks
where gas was searched for and not found,
without violating observations \citep{hillenbrand-2008}.
If present, this hardly detectable gas would heavily affect the
disk's physics and evolution and could necessitate revisions to standard
theories of disk evolution and planet formation.

The goal of our work is to analyze the effects of gas
on the dynamical evolution of solids.
We would like to find out whether gas, if present in larger amounts
or with a different radial distribution than usually assumed, would alter
the dust distribution and thus the brightness profile of a debris disk
in such a way as to show up in the observations.
We follow the approach first suggested by \citet{thebault-augereau-2005}
who applied it to the the $\beta$~Pic system: we first postulate
a certain amount and spatial distribution of gas in one or another
debris disk system, then compute a steady-state distribution of dust
in it, calculate the observables such as brightness profile,
and compare them with available observations.

In Sect.~2 we select and analyze three young debris disk systems
relevant for this study.
Sect.~3 lays down basic theory of the dust production and dynamical
evolution in a debris disk with a gas component.
Sect.~4 describes numerical simulations and Sect.~5 their results.
In Sect.~6 we devise an analytical model and use it to interpret
the numerical results. Sect.~7 contains our conclusions.

%-----------------------------------------------------------------------
\section{Systems}

\subsection{Selection criteria}

We wish to choose several young debris disks in which the
presence of gas in little to moderate amounts has been reported.
Ideally, these should be edge-on systems, so that better constraints on gas
are available from the presence or absence of absorption lines, not
just CO mm emission.
We need dust disks that are spatially resolved, preferably
in scattered light, so that the radial profile of brightness is known.
The age of the disks should not be very
far from the boundary that separates gas- and dust-rich, optically
thick protoplanetary disks from nearly gasless, optically thin debris
disks, which is believed to lie at $\approx 10$~Myr.
The best ages would thus be 10--30~Myr.

We find three systems to satisfy these criteria the best:
$\beta$~Pic, HD~32297, and AU~Mic.
Known facts and key parameters of these systems relevant to our study
are presented in the subsequent sections. We stress, however,
that all three systems should be regarded as ``typical'' examples of
their classes and might be used as a proxy for other systems.
Thus our results, being of interest in the context of particular
objects, could at the same time be considered as generic.

%------------
\subsection{ $\beta$~Pic}

\mbox{}

\vspace*{-\parindent}
{\em Star.}
A 12~Myr-old \citep{zuckerman-et-al-2001} A5V star at $d = 19.44 \pm 0.05$~pc.

{\em Dust and parent bodies.}
The debris disk was first resolved by \citet{smith-terrile-1984}
and later at various  wavelengths \citep[][and references therein]{artymowicz-2000}.
According to \citet{artymowicz-clampin-1997},
the vertical optical depth of the dust disk has a maximum
of $1.53 \times 10^{-2}$ at $60$\,AU,
and has a slope of $-1.7$...$-2$ in the outer part.
\citet{mouillet-et-al-1997} give
$5 \times 10^{-3}$ at 100\,AU with an outer slope of $-1.7$.
Dust mass is roughly $0.05$...$0.5 M_\oplus$ \citep{artymowicz-2000,thebault-augereau-2005},
with $0.1 M_\oplus$ being probably the best estimate 
\citep{zuckerman-becklin-1993,lagrange-et-al-2000}.
Analysis by \citet{augereau-et-al-2001} (their Fig.~1) show that an extended dust
disk produced by the planetesimal belt (``birth ring'') between $80$--$120$\,AU 
would closely match the resolved scattered light images together with the
long-wavelength photometric data.
The dust distribution itself is given in Fig.~2 of \citet{augereau-et-al-2001}.
The radial $\mathit{SB}$ profile of the mid-plane scattered-light images shows
a slope of $-3$...$-4$ outside $130$--$260$\,AU \citep{golimowski-et-al-2006}.
The dust disk around $\beta$~Pic is famous for its large scale asymmetries, which
might be caused by a sub-stellar companion in
the disk \citep[see, e.g.][]{mouillet-et-al-1997,augereau-et-al-2001}.

{\em Gas.}
Before $\beta$\,Pic was known to harbor a debris disk, it was
classified as a shell star by \citet{slettebak-1975} due to its prominent
Ca\,II H \& K absorption. Gas was then re-discovered in absorption by
\citet{hobbs-et-al-1985} and in spatially resolved Na\,I emission by 
\citet{olofsson-et-al-2001}. The atomic hydrogen content of the disk
was constrained by \citet{freudling-et-al-1995} to be $< 2$\,$M_{\oplus}$,
and the molecular column density to be
$< 3 \times 10^{18}$\,cm$^{-2}$ by
\citet{desetangs-et-al-2001}, which corresponds to $\lesssim 0.2$\,M$_{\oplus}$,
assuming the gas to be distributed in the disk \citep{brandeker-et-al-2004}. 
\citet{brandeker-et-al-2004} observed spatially extended gas emission from
a number of elements (including Na\,I, Fe\,I, and Ca\,II), and derived a
spatial distribution for the gas
\be
 n(H) = n_0
      \left[
        \left( r \over r_0 \right)^{2.4}
        +
        \left( r \over r_0 \right)^{5.3}
      \right]^{-1/2}
\label{bpic_gas_profile}
\ee
with
$n_0 = 2.25 \times 10^3 \cm^{-3}$ for an assumed solar composition,
and $r_0 = 117$\,AU, leading to a total gas mass of $0.1\,M_\oplus$.
They also investigated a metal-depleted case $n_0  = 10^6 \cm^{-3}$ 
(implying a total gas mass of $40\,M_\oplus$), which they found to be
in contradiction with the observed upper limits on hydrogen.

%------------
\subsection{HD 32297}

\mbox{}

\vspace*{-\parindent}
{\em Star.}
A 30~Myr-old A5V star \citep{maness-et-al-2008} at $d = 113 \pm 12$~pc. 

{\em Dust and parent bodies.}
The dust disk was first resolved with HST/NICMOS in scattered light by \citet{schneider-et-al-2005}
up to 400\,AU.
The surface brightness ($\mathit{SB}$) of the SW wing is fitted
by a power law with index $-3.6$, while the NE side shows a break at 200\,AU:
the inner part has a slope of $-3.7$, whereas the outer one $-2.7$.
\citet{kalas-2005} resolved the disk in the R-band between 560 and 1680\,AU.
The mid-plane slopes were found to be $-2.7$ and $-3.1$ for NE and SW wings,
respectively, with strong asymmetries.
\citet{moerchen-et-al-2007b} resolved the disk with Gemini South/T-ReCS
in thermal emission at 12 and $18\mum$ up to 150\,AU.
Resolved images by \citet{fitzgerald-et-al-2007} taken with Gemini North at $11\mum$
revealed a bilobed structure with peaks at $\sim 65$\,AU from the star.
\citet{maness-et-al-2008} marginally resolved the disk with CARMA at 1.3~mm.

The spectral energy distribution (SED) fitting by \citet{fitzgerald-et-al-2007} suggests
a population of larger grains~-- and therefore a location of the birth
ring~-- at $\approx 70$--$80$\,AU.
The vertical optical depth of the dust disk at the same distance is
$4 \times 10^{-3}$ \citep{maness-et-al-2008}.
1.3 mm measurements by \citet{maness-et-al-2008} point to the
existence of a third population of even larger grains at a characteristic
stellar distance of 50\,AU, which probably comprises $\ge 95$~\% of the
total dust mass.
Dust mass required to fit the SED up to far-infrared wavelengths
is roughly $0.02\,M_\oplus$,
but 1.3\,mm flux may require as much as $1\,M_\oplus$ of dust
\citep{maness-et-al-2008}.

{\em Gas.}
\citet{redfield-2007} found an intriguingly strong Na\,I absorption.
Assuming the morphology and abundances of the stable gas component to be
the same as for $\beta$~Pic, and that the gas disk extends up to 1680\,AU
as debris disk does, he derived the total gas mass of 
$\sim 0.3\,M_\oplus$.
The absence of the observable CO $J=2-1$ emission with CARMA places an 
upper limit on the gas mass of $\sim 100\,M_\oplus$ \citep{maness-et-al-2008}.

%------------
\subsection{AU Mic}

\mbox{}

\vspace*{-\parindent}
{\em Star.}
A 12 Myr-old dM1e flare star, a member of the $\beta$~Pictoris Moving Group
at $d = 9.94 \pm 0.13$\,pc.
It is the closest known debris disk resolved in scattered light.

{\em Dust and parent bodies.}
The debris disk was first resolved in R-band  by \citet{kalas-et-al-2004}
and \citet{liu-2004}.
Later on, it was resolved with HST/ACS by \citet{krist-et-al-2005} and
in the H-band with Keck AO by \citet{metchev-et-al-2005}.
The dust fractional luminosity is $6 \times 10^{-4}$ \citep{liu-2004}.
The dust mass (up to 1~mm) is estimated to be $\sim 2\times 10^{-4} M_\oplus$
\citep{augereau-beust-2006}, but sub-mm fluxes require
$1 \times 10^{-2} M_\oplus$ \citep{liu-et-al-2004}.
The birth ring of planetesimals is believed to be located at 35\,AU \citep{augereau-beust-2006}.
An R-band $\mathit{SB}$ profile slope of $-3.8$ between 35--200\,AU
was found by \citet{kalas-et-al-2004},
whereas \citet{liu-2004}, \citet{krist-et-al-2005}, and \citet{fitzgerald-et-al-2007b}
derived ${\mathit SB}$ slopes in the range $-3.8$...$-4.7$.
Like $\beta$\,Pic and HR 32297, the disk of AU~Mic possesses
asymmetries, which are probably formed by the dynamical influence of 
planets \citep{liu-2004}.

{\em Gas.}
Non-stringent upper limits on the gas mass were found from
non-detection of CO 3-2 emission by \citet{liu-et-al-2004} ($< 1.3\,M_\oplus$)
and $H_2$ UV absorption by \citet{roberge-et-al-2005} ($< 7 \times 10^{-2}\,M_\oplus$).
\citet{france-et-al-2007} tentatively detected and analyzed fluorescent $H_2$ emission.
Within the observational uncertainties, the data are consistent with gas
residing in the debris disk, although other possibilities such as a cloud that
extends beyond the disk cannot be completely ruled out.
They found a very low total gas mass between
$\sim 4 \times 10^{-4}\,M_\oplus$ and $\sim 6 \times 10^{-6}\,M_\oplus$,
consistent with upper limits $\lesssim 10^{-4}$\,$M_\oplus$ obtained from a search 
for optical absorption lines from Ca\,I, Ca\,II and Fe\,I by
\citet{brandeker-jayawardhana-2008}.

%-----------------------------------------------------------------------

\section{Basic theory}

\subsection{General picture}

Throughout this paper, we adopt the following standard scenario of a debris disk
evolution
\citep[e.g.][]{krivov-et-al-2006,strubbe-chiang-2006,%
thebault-augereau-2007,krivov-et-al-2008,thebault-wu-2008}:
\begin{itemize}
\item
There is a relatively narrow belt of planetesimals (``birth ring'')
in orbits with moderate eccentricities and inclinations.
We assume that this birth ring is located where the scattered image brightness peaks.
Note that the systems resolved at (sub)-mm wavelengths usually exhibit a bright
ring of approximately the same radius. 
\item
Orbiting planetesimals in the birth ring undergo collisional cascade that
grinds the solids down to dust.
We assume that the dust grains with radii $[s, s+ds]$ are produced in the birth
ring at a constant rate $\dot{N} ds$, where
\be
  \dot{N} \propto s^{-q} .
\label{Ndot}
\ee
The parameter $q$ is unknown. However, a usual assumption~--- which we will follow
unless stated otherwise~-- is $q=3.5$.
\item
At smallest dust sizes, stellar radiation pressure effectively reduces the mass
of the central star and quickly (on the dynamical timescale)
sends the grains into more eccentric orbits, with
their pericenters still residing within the birth ring
while the apocenters are located outside the ring.
As a result, the dust disk spreads outward from the planetesimal belt.
The smaller the grains, the more extended their ``partial'' disk.
\item
The dust grain orbits undergo slower modifications due to gas drag
and experience gradual loss due to mutual collisions.
\end{itemize}

\subsection{Stellar gravity and radiation pressure}

We require that the disk is optically thin, so that each dust grain is fully
exposed to stellar radiation at any location in the disk.
Since the radiation pressure is proportional to $r^{-2}$, as is the stellar gravity,
a dust grain experiences ``photogravity'', i.e. gravity of a
star with an ``effective stellar mass'' $M_\mathrm{eff}$:
\be
M_\mathrm{eff} = M_\star \, (1 - \beta) \, ,
\ee
where $\beta$ is the ratio of radiation pressure to gravity
\citep{burns-et-al-1979}:
\be
\nonumber
\beta
= 0.5738 \; Q_{\mathrm{pr}}
  \left(  1 \g \cm^{-3} \over \rho_\mathrm{bulk} \right)
  \left(  1 \mum \over s \right)
  {L_\star / L_\odot \over M_\star / M_\odot} .
\label{beta}
\ee
Here, $Q_{\mathrm{pr}}$ is the radiation pressure efficiency (henceforth set to unity),
$\rho_\mathrm{bulk}$ the material density of particles (we set it to $3.3\g\cm^{-3}$),
$s$ their radius, and $L_\star$ and $M_\star$ are stellar luminosity
and mass, respectively.

\subsection{Gas drag}

We assume that the gas distribution remains unaffected by that of dust and
that gas simply exerts a drag force on the dust particles.
If the gas mass is larger than the dust mass, this assumption is natural.
In the case where both are comparable,
the validity of this assumption will be checked later a posteriori.
Indeed, we will choose the same initial distributions for dust and gas
and will see that these will not diverge considerably in the course
of the disk evolution.

Gas orbits the star at a sub-keplerian speed
\be
v_\mathrm{g} = v_\mathrm{K} \, \sqrt{1 - \eta} \, ,
\ee
where $v_\mathrm{K}$ is the circular Keplerian velocity
and $\eta$ is the ratio of the force that supports gas against
gravity to the gravity force.

There exist two possible reasons for sub-keplerian rotation of a gas disk.
One is the case of a thermally-supported disk, in which the
sub-keplerian rotation stems from the gas pressure gradient.
However, around $\beta$~Pic the gas at non-solar composition has been observed,
which is dominantly supported by radiation pressure rather than gas pressure
\citep{fernandez-et-al-2006,roberge-et-al-2006}.
Thus we also consider another case, where gas is supported against stellar gravity
by radiation pressure.

For a thermally-supported gas disk, we follow
a standard description of the dust aerodynamics
\citep{weidenschilling-1977},
generalized to the presence of radiation pressure
\citep{takeuchi-artymowicz-2001,thebault-augereau-2005,herrmann-krivov-2007}.
The factor $\eta$ is
\be
 \eta =  - (GM_\star)^{-1} {r^2 \over \rho_\mathrm{g}} {d P \over d r} .
\label{eta def}
\ee
Here, $GM_\star$ is the gravitational parameter of the central star
and $P$ the gas pressure,  proportional to gas density $\rho_\mathrm{g}$ and
gas temperature $T_\mathrm{g}$:
\be
 P
 = 
 \rho_\mathrm{g} {k T_\mathrm{g} \over \mu_\mathrm{g} m_\mathrm{H}}
\ee
with the Boltzmann constant $k$,
the mean molecular weight $\mu_\mathrm{g}$ and
the mass of a hydrogen atom $m_\mathrm{H}$.

The gas temperature and density are usually taken to be power laws
\be
 T_\mathrm{g} \propto  r^{-p}
 \label{T_g}
\ee
and
\be
 \rho_\mathrm{g} \propto  r^{-\xi}.
 \label{xi}
\ee
Then, Eq.~(\ref{eta def}) takes the form
\be
 \eta
 =
 (p+\xi)
 {k T_\mathrm{g}^0 \over \mu_\mathrm{g} m_\mathrm{H}}
 {r_0 \over GM_\star}
 \left( r \over r_0 \right)^{1-p} ,
\label{eta}
\ee
where $T_\mathrm{g}^0$ is the gas temperature at a reference distance $r_0$.
The $\eta$ ratio only depends on the gas temperature slope $p$ and its
density slope $\xi$, but not on the gas density at a given distance.
(Note that sub-keplerianity requires $p+\xi > 0$.)
As $T_\mathrm{g}^0 \propto L_\star^{1/4}$,
the dependence on the stellar parameters is weak:
$\eta \propto L_\star^{1/4}/M_\star$.

We now consider the case where gas is supported against stellar gravity by
radiation pressure. 
Denoting by $\beta_{\mathrm{gas}}$ the effective radiation pressure coefficient acting on the gas,
its speed is 
\be
v_\mathrm{g} = v_\mathrm{K} \, \sqrt{1 - \beta_{\mathrm{gas}}} \, ,
\ee
yielding a simple relation
\be
\eta = \beta_{\mathrm{gas}} .
\label{eta rad}
\ee

\bigskip
Regardless of the mechanism that supports the gas disk against gravity,
the gas drag force on a dust grain is expressed by
\citep{takeuchi-artymowicz-2001}
\be
  \vecbold{F}_\mathrm{D}
  =
  - \pi \rho_\mathrm{g} s^2
    \left(v_\mathrm{T}^2+ \Delta v^2\right)^{1/2}
    \Delta \vecbold{v} \, ,
\ee
which combines the subsonic and supersonic regimes.
Here, $\Delta \vecbold{v} \equiv \vecbold{v}_\mathrm{d} - \vecbold{v}_\mathrm{g}$
is the difference between the dust velocity  $\vecbold{v}_\mathrm{d}$
and the gas velocity  $\vecbold{v}_\mathrm{g}$, and
$v_\mathrm{T}$ is the gas thermal velocity:
\be
  v_\mathrm{T}
  = 
  {4 \over 3}
  \left( 8 k T_\mathrm{g} \over \pi \mu_\mathrm{g} m_\mathrm{H} \right)^{1/2} .
\ee
For later discussions of the timescales, we also define the stopping time,
\be
  T_{\mathrm{stop}}
  =
  {4 \over 3}
  {\rho_\mathrm{d} \over \rho_\mathrm{g} }
  {s \over v_\mathrm{T}}
  {1 \over \sqrt{1 + \Delta v^2 / v_\mathrm{T}^2} } ,
\label{T_stop}
\ee
the time interval over which $\Delta \vecbold{v}$ would be reduced by a factor of $e$
if the drag force were constant.

\subsection{Gas temperature}
\label{sect:gastemp}
As we saw in the previous section, dust dynamics is expected to depend sensitively on
the gas temperature, in particular on its radial gradient $p$.
A commonly used assumption is that the gas shares the dust temperature profile
\citep{kamp-vanzadelhoff-2001}, which in the simple blackbody approximation gives
$p=1/2$. This is a reasonable assumption if gas-dust interaction is strong and
the photo-electric heating weak, but may not be valid in general. Indeed,
in case of strong UV environments the photo-electric effect on dust can be the
dominant heating source of the gas and lead to a dust drift instability  
\citep{klahr-lin-2005,besla-wu-2007}. 
For sufficiently high dust content, one may expect (at every distance)
a power-law relation between the gas temperature and the number density
of dust.
Following \citet{klahr-lin-2005}, Eq.~(\ref{T_g}) generalizes to
\be
 T_\mathrm{g} \propto  \rho_\mathrm{dust}^\gamma \; r^{-p} 
\label{gamma}
\ee
with $\gamma \ge 0$. A first-order heating-cooling analysis suggests
$\gamma \sim 1$ as a typical value \cite[see the Appendix of][]{klahr-lin-2005}.

\begin{figure}[bth!]
  \centerline{\includegraphics[width=0.45\textwidth]{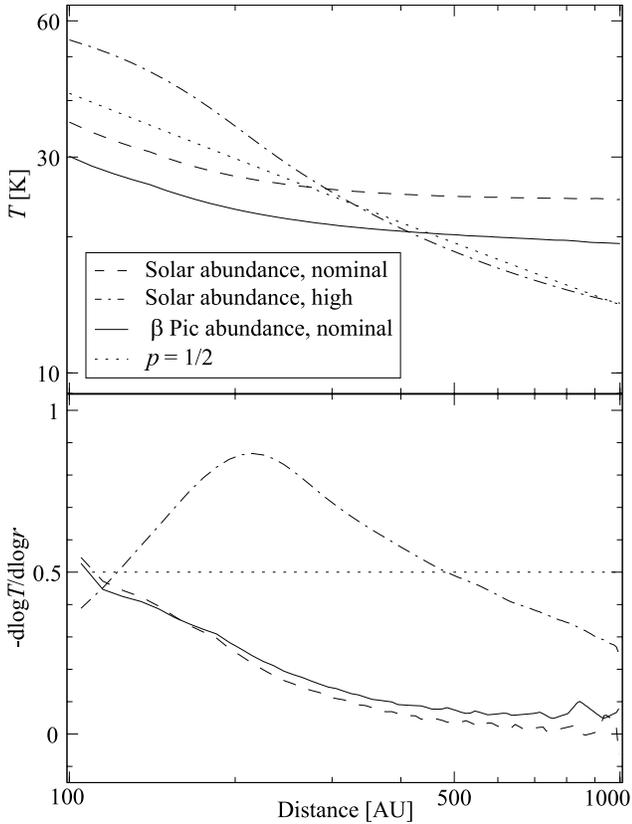}}
  \caption{
  Top: Gas temperature in the $\beta$~Pic disk
  as a function of distance from the star, for
  three different assumptions about gas composition and density.
  (Dashed, dash-dotted, and solid lines correspond to
  tns, ths, and rns models, respectively, of Table~\ref{tab_systems} below.)
  Bottom: Radial slope $p$ of the gas temperature.
  \label{fig_T_gas}
  }
\end{figure}

To evaluate how valid the $p=1/2$ assumption is for a more detailed
model of the thermal balance, we used the 
code \textsc{ontario}
(Zagorovsky, Brandeker, \& Wu, in prep.).
\textsc{ontario} is tuned to model gas emission during the debris disk
phase, given input parameters related to the gas/dust disk structure,
elemental abundances, and the stellar luminosity spectrum.
\textsc{ontario} computes the ionization and thermal balance
self-consistently, with particular care taken of heating/cooling
mechanisms (the most important being photo-electric and ionization
heating, and cooling by C\,II $158\mum$). The major simplifying
assumptions are that the gas is considered to be in atomic/ion form (no
molecules), and that the disks are optically thin (i.e.\ no chemistry
and simplified radiative transfer), conditions that are expected to be
closely met by debris disks around A and F stars, but not necessarily around
stars of later spectral type.
Using the same gas and dust profiles as in the
dynamical simulations (i.e., surface density slope of -1.5 corresponding to
a mid-plane density slope of -2.5), we computed the mid-plane temperature
for three different cases, as shown in Fig.~\ref{fig_T_gas} (top):
\begin{enumerate}
\item
  an $M_{\mathrm{gas}} = 0.1\,M_{\oplus}$ 
  disk at solar abundance with hydrogen entirely in atomic form (giving the mean
  molecular weight $\mu_\mathrm{g} = 1.3$)
\item
  a similar model but with $M_{\mathrm{gas}} = 10\,M_{\oplus}$
\item
  an $M_{\mathrm{gas}} = 0.1\,M_{\oplus}$ model with 
  $\beta$\,Pic abundances (i.e.\ solar abundance except 20$\times$ carbon, no helium, 
  and $10^{-3}\times$ hydrogen, giving $\mu_\mathrm{g} = 11.3$, as motivated by the 
  inventory of gas observed around $\beta$\,Pic, compiled by \citealt{roberge-et-al-2006}).
\end{enumerate}
For comparison, the $T_\mathrm{g} \propto r^{-1/2}$ is overplotted
(with arbitrary normalization).
The bottom panel of Fig.~\ref{fig_T_gas} shows the local $p = -d\log T /d\log r$ exponent, 
which can be compared to the $p=1/2$ value. The numerical noise seen in the plot is due to the 
limited precision
of the code ($dT \sim$ 0.1\,K). The temperature is not a power law, but
varies smoothly with a local exponent that is constrained to the interval $-1$...$0$;
a $p=1/2$ power law therefore seems to be a reasonable approximation.

\subsection{Radial dynamics of dust under photogravity and gas drag}
\label{sect_rad_dyn}

We start with a thermally-supported gas disk.
Since gas adds or removes angular momentum to or from solid particles, it causes
them to spiral outward or inward, until a certain size-dependent stability distance is reached,
at which the gas pressure gradient and
the stellar radiation pressure balance each other
\citep{takeuchi-artymowicz-2001}:
\be
\beta(s) = \eta(s, r_\mathrm{stab}) ,
\label{stability}
\ee
which can be solved for $r_\mathrm{stab}$ for a given $s$ or vice versa.
Note that additional radial forces such as photophoresis \citep{herrmann-krivov-2007}
can be included by simply adding them to the left
part of Eq.~(\ref{stability}).

We now turn to the case where the gas is radiation pressure-supported.
Then, $\eta$ is independent
of distance, so that there is only one value for
$\eta$ over the entire disk and for all particle sizes.
In this case, the particles with $\beta < \eta$ ($\beta > \eta$)
will spiral inward (outward) from the birth ring;
those that just have $\beta = \eta$ will have no radial
motion and will stay in the parent belt.
Therefore, only grains with $\beta > \eta$ will make a contribution
to the outer disk. They all will be drifting outward all the way through
in a steady-state regime.

\subsection{Vertical dynamics of dust under photogravity and gas drag}

Apart from the radial dynamics described above, dust grain orbits also have
vertical evolution. The essential effect is dust sedimentation (or settling)
toward the mid-plane of the disk. It happens on a timescale $T_{\mathrm{sett}}$
that depends on the particle size and the vertical profile of the gas density.
Generally, $T_{\mathrm{sett}}$ also changes with time because
of the grain's radial drift.
If, for dust grains that contribute to visible brightness,
the settling timescale is comparable to the stopping time
($T_{\mathrm{sett}} \sim T_{\mathrm{stop}}$), a combination of settling and radial drift
would cause the aspect ratio of the outer disk to decrease with distance
(``anti-flaring''). Thus in principle, a comparison of the observed
vertical distribution of brightness in an edge-on debris disk with
a modeled one may offer another method of constraining the gas densities.
In the present paper, however, we confine our analysis to the radial distribution
and radial brightness profiles.

\subsection{Collisions}
Collisions are the main mechanism that limits the lifetime of dust grains
in the outer disk, if they are large enough not to be blown away by radiation
pressure or rapidly dragged away by gas.

Collisional outcome is known to depend sensitively upon the relative velocities.
Both projectiles are disrupted if their relative velocity $v_\mathrm{rel}$ exceeds
\citep[see, e.g.][their Eq. 5.2]{krivov-et-al-2005}
\be
  v_{\mathrm{cr}} = \sqrt{
    {2 (m_\mathrm{t} + m_\mathrm{p})^2  \over m_\mathrm{t} m_\mathrm{p}}  Q_D^\star
                         } ,
\label{disruption}
\ee
where
$m_\mathrm{t}$ and $m_\mathrm{p}$ are masses of the two colliders and
$Q_D^\star$ is the critical energy for fragmentation and
dispersal, which is $ \sim 10^8\erg\g^{-1}$ at dust sizes \citep[e.g.][]{benz-asphaug-1999}.
If $v_\mathrm{rel} < v_{\mathrm{cr}}$, the collision may result in partial fragmentation (cratering),
restitution or, at low velocities, merging of the two colliders.

The relative velocity $v_\mathrm{rel}$ mainly stems
from the difference in the radial velocities of different-sized grains $|v_r|$,
as set by stellar photogravity and gas drag.
In Sect.~5, we will see that $v_\mathrm{rel}$ is typically high enough for
catastrophic collisions to occur even if the gas density is high.

%-----------------------------------------------------------------------
\section{Numerical simulations}

\subsection{Setups for simulations}

For each of the three systems, we adopt a fixed set of 
parameters for the central star and solids (planetesimals and dust)
and test various gas density models (Table~\ref{tab_systems}).
One model may differ from another in four respects:

\newcommand{\tb}[1]{\makebox[6mm][c]{#1}}

\begin{table*}[t]
\begin{center}
  \caption{Models for $\beta$~Pic, HD~32297, and AU Mic.
  \label{tab_systems}
  }
  \begin{tabular}{l|cccccccccc|cc|cc}
  \hline
			& \multicolumn{10}{|c|}{\makebox[3cm]{$\beta$~Pic}} & \multicolumn{2}{|c|}{\makebox[1.2cm]{HD~32297}} & \multicolumn{2}{|c}{\makebox[1.2cm]{AU~Mic}}\\
  \hline
  \multicolumn{15}{l}{\em Star} 			\\
  Luminosity, $L_\odot$	& \multicolumn{10}{|c|}{$8.70$}		            & \multicolumn{2}{|c|}{$5.40$}			& \multicolumn{2}{|c}{$.092$}\\
  Mass, $M_\odot$	& \multicolumn{10}{|c|}{$1.75$}		            & \multicolumn{2}{|c|}{$1.80$}			& \multicolumn{2}{|c}{$.50$}\\
  Age, Myr		& \multicolumn{10}{|c|}{$12$}		            & \multicolumn{2}{|c|}{$30$}			& \multicolumn{2}{|c}{$12$}\\

  \hline
  \multicolumn{14}{l}{\em Planetesimals and dust}	\\
  Birth ring, AU	& \multicolumn{10}{|c|}{110-130}		    & \multicolumn{2}{|c|}{60--80}			& \multicolumn{2}{|c}{30--40}\\
  Max optical depth     & \multicolumn{10}{|c|}{$5 \times 10^{-3}$}         & \multicolumn{2}{|c|}{$4 \times 10^{-3}$}		& \multicolumn{2}{|c}{$6 \times 10^{-4}$}\\
  Init. surf. density slope & \multicolumn{10}{|c|}{$-1.5$}		            & \multicolumn{2}{|c|}{$-1.5$}			& \multicolumn{2}{|c}{$-1.5$}\\
  $[s_{min}, s_{max}]$, $\mum$ & \multicolumn{10}{|c|}{$[1,    86]$}        & \multicolumn{2}{|c|}{$[0.6,  52]$}                & \multicolumn{2}{|c}{$[0.04, 3.2]$}\\

  \hline
  \multicolumn{15}{l}{\em Gas} 			\\
  Model identifier	  &\tb{000}&\tb{tns}&\tb{t0ns}&\tb{t1ns}&\tb{rns}&\tb{r0ns}&\tb{r1ns}&\tb{ths}&\tb{thf}&\tb{tvs}	&\tb{tns}&\tb{ths}	&\tb{tns}             &\tb{ths}    \\
  t- or r-support         & -      & t 	    & t       & t       & r      & r       & r       & t      & t      & t              & t      & t            & t                   & t          \\
  Temperature slope $p$   & -      & $1/2$  & $0$     & $1$     & $1/2$  & $0$     & $1$     & $1/2$  & $1/2$  & $1/2$          & $1/2$  & $1/2$        & $1/2$               & $1/2$      \\
  Mass, $M_\oplus$	  & -      & $.05$  & $.05$   & $.05$   & $.05$  & $.05$   & $.05$   &  5     &  5     &  50           	& .15    &  15   	& $2\times 10^{-4}$   & $2\times 10^{-2}$ \\
  Surface density slope   & -      & -1.5   & -1.5    & -1.5    & -1.5   & -1.5    & -1.5    & -1.5   & -1.0   & -1.5 		& -1.5 	 & -1.5 	& -1.5                & -1.5             \\
  Outer edge, AU$^\star$  & \multicolumn{10}{|c|}{$600$}                       & \multicolumn{2}{|c|}{$350$}       & \multicolumn{2}{|c}{$175$}\\
  \hline
  \end{tabular}
\end{center}
\smallskip
{\small $^\star$ The arbitrary truncation distance (we set it to five times the central distance of the respective birth ring).
It only affects the conversion of total gas mass to gas density.
In that conversion, we also use a constant semi-opening angle of $15^\circ$ for all disks, assuming the density to be
constant with height.}
\end{table*}

\begin{enumerate}

\item
{\em Gas support mechanism.}
We consider two possible reasons for sub-keplerian rotation
of a gas disk: a thermally-supported disk,
which seems reasonable in the
case of primordial gas at solar composition,
and a radiation pressure-supported disk,
which could be more appropriate for secondary gas with
a non-solar composition.
\citet{fernandez-et-al-2006} computed the effective radiation pressure on the
ionized gas to be $\beta \sim 4$ in the $\beta$~Pic disk (see their Fig.~4), 
assuming solar abundances. Since the mass is dominated by (inert) carbon, which was later 
found to be overabundant by 
a factor 20 \citep{roberge-et-al-2006}, the effective radiation pressure coefficient
acting on the gas including 20$\times$ carbon is estimated to be
$\beta_{\mathrm{gas}} \sim 4/20 = 0.2$. Thus for the radiation pressure-supported
disks we set
$\eta = \beta_{\mathrm{gas}} = 0.2$.

\item
{\em Radial slope of the gas temperature}.
In most of the models, we assume $p=1/2$, but also test
$p=0$ and $p=1$ to bracket the behavior of expected
temperature profiles (Sect.~\ref{sect:gastemp}).

\item
{\em Total amount of gas}.
In the nominal gas models, the total gas mass is taken as retrieved
from the gas observations:
$0.1 M_\oplus$ for $\beta$~Pic,
$0.3 M_\oplus$ for HD~32297,
$4 \times 10^{-4} M_\oplus$ for AU~Mic.
As we consider here only the outer part of the disks, outside the birth
ring, we simply halve these masses. Indeed, in the $\beta$~Pic gas disk,
the masses of the inner and outer disks are nearly equal
(two terms in Eqs. (4) or (5) of \citet{brandeker-et-al-2004}
make comparable contributions).
Interestingly, the above gas masses are roughly comparable with the dust mass
(gas:dust ratio $\approx$ 1:1).
However, as explained in the introduction, the systems may contain much more
gas than is evident in the observations.
For this reason, we consider
high gas mass models, in which the total gas mass is 100 times the nominal mass.
With this choice, the gas-to-dust ratio is a standard 100:1.
Finally, we try a very high gas case, in which the gas mass is
ten times higher than in the high gas models.

\item
{\em Radial slope of the gas surface density.}
If the gas is secondary, one could expect the gas profile to approximately
follow the dust profile \citep[e.g.,][]{czechowski-mann-2007}.
For our three systems, the latter falls off as $\propto r^{-1}$...$\propto r^{-2}$.
Slopes in this range are also expected on theoretical grounds
in a standard  ``birth ring - collisionally evolving gas-free disk'' model
explained in Sect. 3.1.
Thus our standard choice is to set the surface density of both gas and dust
to be $\propto r^{-1.5}$
initially.
However, if the gas is primordial, i.e. a remnant of an accretion disk,
the profile is more uncertain.
One could still expect $\propto r^{-1.5}$ (consistent with an isothermal
steady-state solution for a viscous accretion disk), but, for instance,
a flat density profile $\propto r^{-1}$ (another known steady-state solution)
could also be possible.
\end{enumerate}

Accordingly, each model has an identifier $XpYZ$, where \\[-5mm]
\begin{itemize}
\item $X$ indicates the type of the gas disk
  \begin{itemize}
   \item 0 (no gas)
   \item t (thermally-supported)
   \item r (radiation pressure-supported)
  \end{itemize}
\item $p$ indicates the slope of gas temperature:
  \begin{itemize}
   \item 0 ($p=0$)
   \item 1 ($p=1$)
   \item nothing ($p=1/2$)
  \end{itemize}
\item $Y$ indicates the total amount of gas:
  \begin{itemize}
   \item 0 (no gas)
   \item n (``nominal'' gas mass)
   \item h (``high'' gas mass)
   \item v (``very high'' gas mass)
  \end{itemize}
\item $Z$ indicates the slope of the gas density profile:
  \begin{itemize}
    \item 0 (no gas)
    \item s (standard, surface density falls off as $1.5$)
    \item f (flat,     surface density falls off as $1.0$)
  \end{itemize}
\end{itemize}

Our ``tpns'' and ``rpns'' models
(a thermally- or radiation pressure-supported gas disk
with various gas temperature profiles, a nominal gas content,
and a standard surface density slope)
are expected to simulate {\em secondary gas} whose production
is somehow related to dust.
In contrast, the ``ths'', ``tvs'', and ``thf'',
models would emulate a possible remnant of {\em primordial gas}
in the outer part of the system.

As reference models, we treat ``tns'' and ``ths'' cases
(a thermally-supported gas disk with a nominal or high gas content
and a standard surface density slope).
We ran these models for all three systems.
In the case of $\beta$~Pic, we additionally tested other
disks, as listed in Tab.~\ref{tab_systems}.

\subsection{Numerical integrations}

We now describe the procedure to compute individual trajectories
of dust grains as well as the overall brightness profiles of the
disks.

First, we assumed that dust parent bodies are orbiting in the ``birth ring''.
Their orbital semimajor axes were uniformly distributed
within the birth ring as specified in Table~\ref{tab_systems}
and their eccentricities between $[0.0, 0.1]$.

In each run with a thermally-supported gas disk,
we launched 500 particles with radii distributed uniformly in a log scale
between $s_{min}$ and $s_{max}$.
The size ranges $[s_{min}, s_{max}]$ used 
correspond to a $\beta$-ratio interval from $0.9$ down to $0.01$.
The minimum $\beta$ value of 0.01 is chosen to
cover all grains which can have their stability distance
in or outside the center of the birth ring
(see Fig.~\ref{fig_r_stab} below).
Upon release, the particles instantaneously acquire
orbits with semimajor axes and eccentricities different from those of their
parent bodies, which is a standard radiation pressure effect \citep{burns-et-al-1979}.

The forces included stellar gravity, direct radiation pressure,
the Poynting-Robertson force, and gas drag.
The drag force due to the stellar wind, which is known to be important for AU Mic 
\citep{strubbe-chiang-2006,augereau-beust-2006},
is not included. The reason is that a nearly radial stellar
wind could no longer exist in the outer disk considered here,
between roughly 35 and 300\,AU, because of the interactions with the
presumed rotating gas.
Indeed, a simple estimate, assuming a mass loss
rate of $\sim 10^{-12} M_\odot \yr^{-1}$ and the stellar wind velocity of
$\sim 400\km\second^{-1}$,
yields the total mass of the stellar wind particles in the outer disk
of $\la 10^{-6} M_\oplus$, about three orders of magnitude less than
the mass of rotating gas in the nominal model.

The particle orbits were followed with the \citet{everhart-1974,everhart-1985}
integrator of 15th order with an adaptive step size.
The integrations ended upon one of the following, whichever was the earliest:
(i) a grain came as close as 10\,AU to the star;
(ii) a grain reached the distance of 1000\,AU;
(iii) after $10^5$~yr of integration. 
Instantaneous positions of particles were stored each 500 years for 
bound grains and each 5 years for unbound ones.
A typical number of instantaneous positions per system and gas model
was $\sim 10^6$.

The setup for the runs with radiation pressure-supported gas disks
was different from what is described above,
because using the same setup would lead to the following problem.
As explained in Sect. 4.3, we normalize the calculated dust density
in such as way as to arrive at the correct maximum geometrical optical depth
$\tau_0$ (Table~\ref{tab_systems}).
In the usual runs for thermally-supported disks, $\tau_0$ in the birth ring
is dominated by the particles with $\beta < 0.2$. But in the ``rns'' run
these drift inward, so that $\tau$ comes from grains with $\beta \ge 0.2$.
These have smaller cross section, and thus their number density turns out
to be two orders of magnitude higher than in standard runs. Thus the
collisional lifetime becomes quite short, $\sim 1$...$10$ years.
For these reasons, the ``rns'', ``r0ns'', and ``r1ns'' models were run with just 15 grains
(instead of 500) over $10^3$ years (instead of $10^5$),
and the recording time step was as small as 0.5 years
for bound grains and 0.005 years for unbound ones.
The minimum $\beta$ value was set to $\eta = 0.2$ (instead of 0.01),
because in the radiation pressure-supported disks only grains with 
$\beta > \eta$ drift outward from the birth ring.

\subsection{Collisional post-processing}
\label{collpost}

The collisions were applied to the numerical integration results
through the following post-processing algorithm:
\begin{enumerate}

\item
The instantaneous positions of grains stored during the numerical integrations
are distributed into two-dimensional size-distance bins,
$(s_i, r_j)$ or $(\beta_i, r_j)$.
The number of occurrences in each bin $N_{\mathrm{bin}}(\beta_i, r_j)$ 
is converted into the absolute number density
$n(\beta_i,r_j)$ from the known normal geometrical optical depth at the birth ring
(Table~\ref{tab_systems}).
Besides, for each bin we calculate the average radial velocity of its grains,
$v_r(\beta_i,r_j)$.

\item
For each $(\beta_i,r_j)$-bin, the collisional lifetime of its particles
is calculated as follows. We consider all bins with various $\beta_k$
at the same distance $r_j$ and check if Eq.(\ref{disruption}) is fulfilled.
The reciprocal of the collisional lifetime is
\bea
  T_{\mathrm{coll}}(\beta_i,r_j)^{-1}
  &=&
  \sum\limits_k
  n(\beta_k,r_j)
\nonumber\\
  &\times&
    \max [ v_r(\beta_i,r_j), v_r(\beta_k,r_j) ]
\nonumber\\
  &\times&
   \sigma (\beta_i,\beta_k) ,
\label{T_coll}
\eea
where $\sigma$ is the collisional cross section and 
summation is performed over those projectiles that satisfy Eq.(\ref{disruption}).

\item
We then go back to the stored numerical integration results and go along
each trajectory again. For a particle with $\beta_i$ at each time step $t_l$,
we calculate a probability that the particle survives destructive collisions over the
current time step,
$p(\beta_i, t_l) = \exp [-\Delta t / T_{\mathrm{coll}}(\beta_i,r_j)]$,
where
$r_j$ is the distance at the moment $t_l$ and 
$\Delta t = t_l - t_{l-1}$ is time step between two successive stored positions
of the particle. Having $p$ for all previous time steps for the
same grain, we determine the probability for that grain to survive
collisions up to the current moment of time:
\be
 P(\beta_i, t_l) = \prod\limits_{m=0}^l p(\beta_i,t_m) .
\ee

\item Steps 1--3 are repeated iteratively.
From now on, when determining the number densities in step 1,
we use $N_{\mathrm{bin}}(\beta_i, r_j) P(\beta_i, t_l)$
instead of just $N_{\mathrm{bin}}(\beta_i, r_j)$.

\end{enumerate}

Note that we do not compute the exact relative velocities between impacting grains,
since this procedure would be too time consuming, but assume that this velocity
is of the order of the $v_r$ of the grain having the highest radial velocity. In doing so,
we implicitly neglect the azimuthal component of the relative velocity, and our procedure
should thus be considered as giving a lower estimate for impact speeds and thus
shattering efficiency of collisions.
The procedure is computationally fast and converges rapidly to a steady state.
In practice, we perform five iterations.

\subsection{Calculation of surface brightness profiles}

From $N_{\mathrm{bin}}(\beta_i, r_j)$ recalculated after several iterations,
the surface brightness of the disk is computed as
\be
  \mathit{SB}(r_j)
  =
  \mathrm{const} \times
  \sum\limits_i
  N_{\mathrm{bin}}(\beta_i, r_j) s_i^{3-q} r_j^{-3}
\ee
(we set $q=3.5$).
Here, the factor
$s^{3-q}$ provides conversion from particle numbers to their cross section area ($s^2$),
accounts for a logarithmic binning of sizes ($dN/d{\ln s} = s dN/ds$),
and includes the size distribution at production ($s^{-q}$, Eq.~\ref{Ndot}).
In turn, the factor $r^{-1}$ is needed because the surface area of a radial annulus
is proportional to $r$, whereas $r^{-2}$ is a conversion from optical depth to the $\mathit{SB}$.

The latter conversion needs to be explained in more detail.
For the sake of simplicity, we assume here gray, isotropic scattering.
In that case, the brightness of the edge-on disk is known to come
mostly from the part of the line of sight closest to the star
\citep[e.g.][]{nakano-1990,thebault-augereau-2007}.
Further, as mentioned above, we assume a non-flared disk,
with the dust scale height being linearly proportional to $r$.
With these assumptions, the standard brightness integral gives
$\mathit{SB}(r) \propto \tau(r) r^{-2}$.
Therefore, if
\be
\tau(r) \propto r^{-\alpha} 
\ee
then
\be
\mathit{SB}(r) \propto r^{-\mu} 
\ee
with
\be
  \mu = \alpha + 2 .
  \label{mu_alpha}
\ee

%-----------------------------------------------------------------------

\section{Results}

In this section, we start with an analysis of typical single-grain
dynamics and then proceed with $\mathit{SB}$ profiles.

\subsection{Radial drift of dust grains}

\begin{figure}[bth!]
  \centerline{\includegraphics[width=0.45\textwidth]{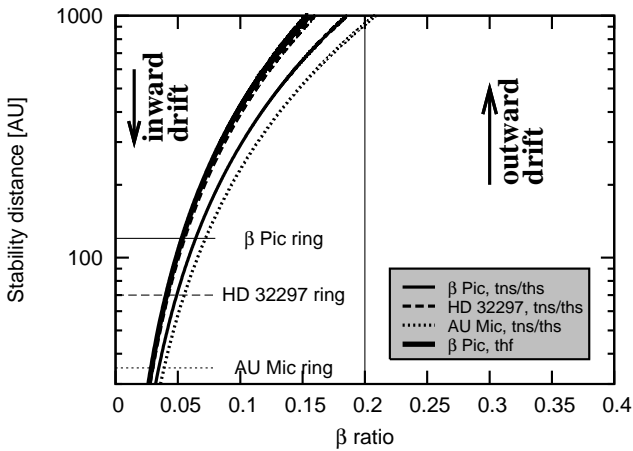}}
  \caption{
    Loci of stability in the $\beta$ ratio -- distance plane
    in thermally-supported gas disks (curves) and
    in a radiation pressure-supported disk (vertical straight line).
    Grains to the left of the stability lines drift inward, those to the right move outward.
    Thin horizontal lines mark the distances at which the birth rings around all three stars are centered.
  \label{fig_r_stab}
  }
\end{figure}

We start with the case of a thermally-supported gas disk.
Numerical solution of Eq.~(\ref{stability}) with $p=1/2$ is shown in Fig.~\ref{fig_r_stab}.
It shows that the stability distance is less than 1000\,AU only for grains
with $\beta \la 0.15$--$0.20$.
Grains with higher $\beta$ ratios are sweeping outward all the way through the disk.

\begin{figure}[bth!]
  \centerline{\includegraphics[width=0.40\textwidth]{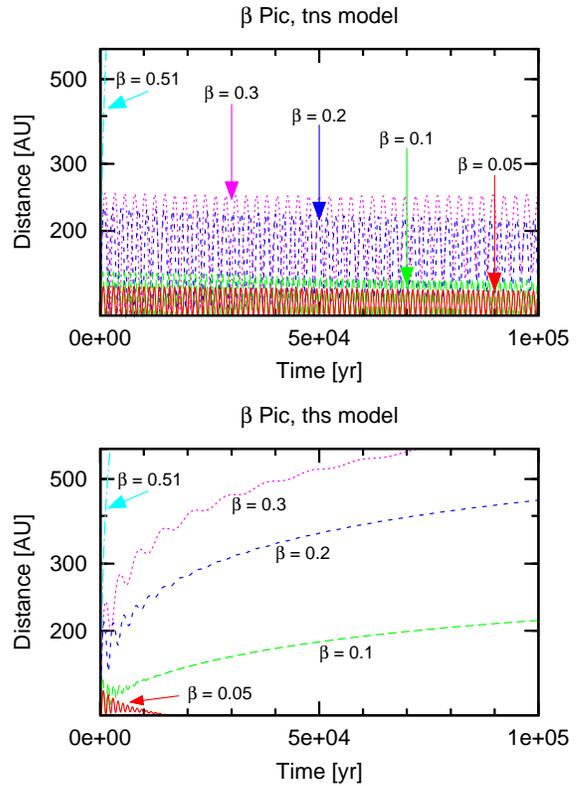}}
  \caption{
   Radial drift of different-size particles in the $\beta$~Pic nominal gas (top)
   and high gas models (bottom).
    \label{fig_r_of_t}
  }
\end{figure}

Fig.~\ref{fig_r_of_t} illustrates how grains with different $\beta$ ratios drift toward
their stability distances in the nominal and high gas $\beta$~Pic models.
It shows in particular how sensitive the drift timescale $T_{\mathrm{drift}}$ (time needed
for a grain to reach its stability distance) is to the the grain sizes and to the gas content.
Further, it tells us that $T_{\mathrm{drift}} \gg T_{\mathrm{stop}}$. One reason for this is that
$T_{\mathrm{stop}}$ increases rapidly with increasing distance. For instance,
in the nominal gas case, the $\beta=0.1$ grains have the stability distance
at $\sim 500$\,AU, but are still at $\sim 150$\,AU after 1~Myr of evolution.
Hence they have $T_{\mathrm{drift}} \gg 1$~Myr, whereas their stopping time
at the birth ring distance is $T_{\mathrm{stop}} \sim 2 \times 10^3$~yr.

We now turn to the case where the gas is radiation pressure-supported.
As explained in Sect.~\ref{sect_rad_dyn},
only the particles with $\beta > \eta$
will spiral outward from the birth ring, and the stability distance does not exist.
Thus the outer disk will only be composed of relatively small grains
in unbound orbits.

\subsection{Radial velocities and collisional outcomes}

\begin{figure*}[bth!]
  \centerline{\includegraphics[width=0.65\textwidth]{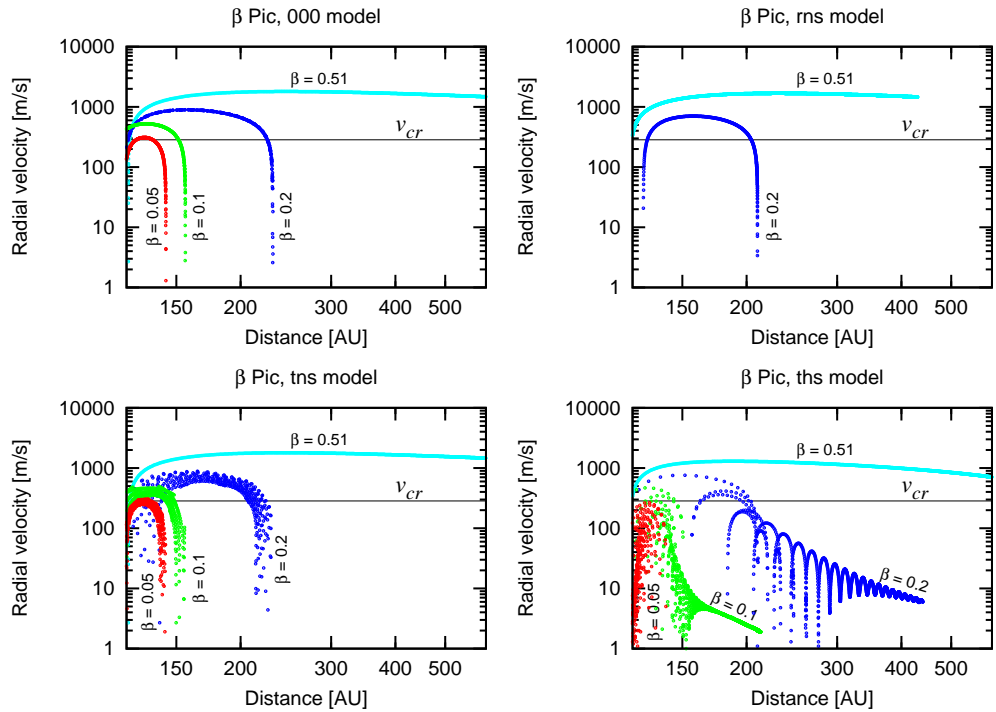}}
  \caption{
   Radial velocity of different-sized grains in the $\beta$~Pic disk
   000 (top left), rns (top right), tns (bottom left) and ths (bottom right) gas models, obtained 
   with numerical 
   integrations described in Sect.~4.
   For comparison, a horizontal straight line gives the minimum relative velocity $v_\mathrm{cr}$
   needed to break up two like-sized colliders
   (this value is independent of sizes, since we assume a constant
   critical energy for fragmentation and dispersal $Q_D^\star = 10^8 \erg\g^{-1}$ for all dust 
   sizes).
  \label{fig_rad_vel}
  }
\end{figure*}

The relative velocity of impacting grains is a crucial parameter, leading
to their destruction if it exceeds $v_\mathrm{cr}$ (Eq.~\ref{disruption}).
As discussed in Sect.~\ref{collpost}, we assume that these impact speeds are of 
the order of particle radial velocities. Figure~\ref{fig_rad_vel} plots $v_r$ for
grains of various sizes, in the $\beta$~Pic disk without gas and for two gas models,
and compares them to the value of $v_\mathrm{cr}\sim 300\m\second^{-1}$ for impacts between
\emph{equal-sized} particles.
The top panel reveals Keplerian U-shape curves, where $v_r>v_\mathrm{cr}$ for all grains
smaller than $s \sim 17 \mum$ ($\beta > 0.05$), except near the apsides.
Under the presence of gas (middle and bottom panels), radial velocities are
damped, and only $\beta \sim 0.1$ ($s \sim 10 \mum$)
grains have $v_r$ exceeding the disruption threshold in the nominal gas case.
For the high gas case, this damping is much more efficient so that after a few
pseudo-orbits, even $\beta \sim 0.2$ ($s \sim 4 \mum$) grains have
$v_r<v_\mathrm{cr}$. The results are similar for the other two systems, i.e., low
$v_r$ for all grains larger than $\beta \sim 0.05$ in the gas free case and
$\beta \sim 0.2$ in the high gas case.

\begin{figure}[bth!]
  \centerline{\includegraphics[width=0.40\textwidth]{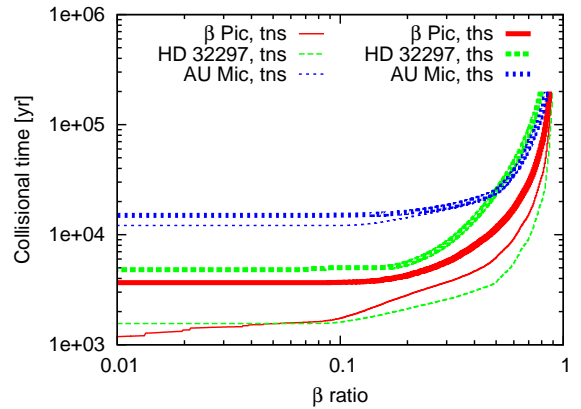}}
  \caption{
   Collisional lifetime of different-sized grains (i.e. those with different $\beta$ ratios),
   obtained in numerical simulations described in Sect.~4.
  \label{fig_T_coll}
  }
\end{figure}

However, collisional disruption plays a significant role in all the systems,
for all gas models considered, and across the whole range of distances in
the outer disk. This is because of the crucial role of collision between
grains of \emph{unequal} sizes. This is easy to understand. Indeed,
Fig.~\ref{fig_rad_vel} shows that very small grains always maintain large radial velocities.
Hence, for a grain of a given size, there is a significant amount
of somewhat smaller grains that are fast enough and still massive enough
to satisfy Eq.~(\ref{disruption}).
To illustrate this, we plot in Figure~\ref{fig_T_coll} the collisional lifetimes of 
different-sized
grains around all three stars, calculated with the algorithm described in section 5.
We see that for $\beta \leq 0.2$, $T_\mathrm{coll}$ is more or less independent of 
particle sizes in all considered cases.
Overall, $T_\mathrm{coll}$ is longer in high-gas models and shorter in nominal-gas ones.
For bound grains ($\beta < 0.5$) in all models, they never exceed $4 \times 10^4$~yr.

\subsection{Surface brightness profiles without gas}

\begin{figure}[htb!]
  \centerline{\includegraphics[width=0.45\textwidth]{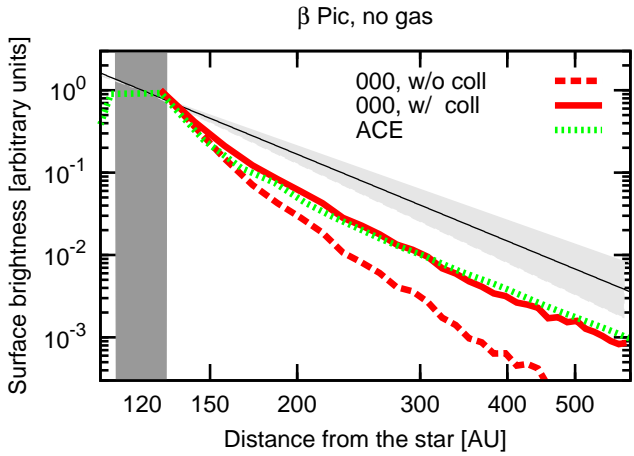}}
  \caption{$\mathit{SB}$ profiles for $\beta$~Pic without gas (thick curves).
  Dashed and solid lines are without and with collisions taken into account, respectively.
  Dotted line is the profile
  obtained with an elaborate collisional code ACE \citep{krivov-et-al-2008}.
  All profiles are normalized to unity at the outer edge of the birth ring.
  Filled vertical bar shows the location of the birth ring.
  The grey-shaded area is bordered by power laws with indices $-3.0$ and $-4.0$,
  and a $-3.5$ slope is marked with a thin straight line.
  \label{fig_sb_profiles_nogas}
  }
\end{figure}

We now turn to $\mathit{SB}$ profiles and start with a disk that does not contain gas
(``000'' model for $\beta$~Pic), both with and without collisions taken into account.
For comparison, we ran our full collisional code ACE \citep{krivov-et-al-2008} that
provides a detailed treatment of collisions with a multitude of collisional outcomes.
The resulting profiles are shown in Fig.~\ref{fig_sb_profiles_nogas}.

Without collisions we get a steep ($\approx -5$) $\mathit{SB}$ profile.
This result is expected when assuming that
grains down to the radiation pressure blow-out limit
($\beta = 0.5$) are produced in the birth ring with a $q=3.5$
size distribution and that the smallest of these grains, which
dominate the outer ring, are then
simply diluted along their eccentric orbits
and thus get underabundant in the birth ring.
\citet{strubbe-chiang-2006} showed that in this case the resulting surface density
profile approximately decreases as $r^{-2.5}$, yielding an $\mathit{SB}$ slope
of $-4.5$ \citep[see][for a more detailed discussion]{thebault-wu-2008}.

In contrast, the $\mathit{SB}$ profile with collisions
turned out to be close to $-3.5$...$-4.0$,
which corresponds to a surface density slope of $\approx -1.5$...$-2.0$.
This slope agrees reasonably well with the one calculated with a full collisional
simulation with ACE. A radial $\mathit{SB}$ slope close to 3.5 is theoretically expected, too.
The difference with the collisionless case is
that the small high-$\beta$ grains
can only be produced and destroyed in the birth ring but spend most of their time in collisionally
inactive regions beyond it.
As a result, their number density follows the $q=3.5$
size distribution {\emph within} the birth ring
(instead of being underabundant as in the collisionless case)
so that the total integrated number of small grains
(taking into account the large fraction outside the birth ring)
is much higher than the one derived from a $q=3.5$ law
\citep[see][]{strubbe-chiang-2006,thebault-wu-2008}.

\begin{figure}[htb!]
  \centerline{\includegraphics[width=0.45\textwidth]{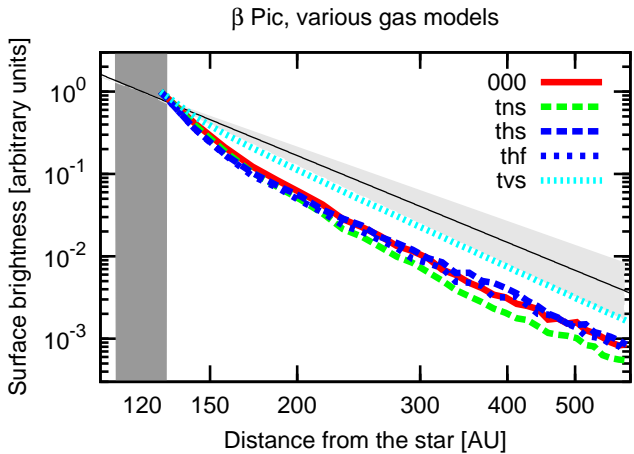}}
  \caption{
  Same as Fig.~\ref{fig_sb_profiles_nogas},
  but for
  several gas models (thick curves).
  Collisions are included.
  \label{fig_sb_profiles_bpic}
  }
\end{figure}

\begin{figure}[htb!]
  \centerline{\includegraphics[width=0.45\textwidth]{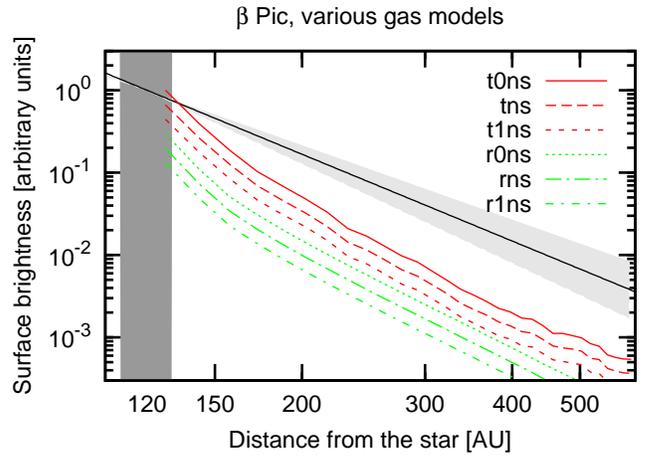}}
  \caption{Same as Fig.~\ref{fig_sb_profiles_bpic}, but for models
  assuming different gas temperature profiles.
  Since all the curves fall almost on top of each other, 
  we artificially lowered each curve listed in the legend
  (starting from ``tns'') relative to the previous one
  by a factor of 1.5.
  \label{fig_sb_profiles_bpic_p}
  }
\end{figure}

\subsection{Surface brightness profiles with gas: $\beta$~Pic}

Fig.~\ref{fig_sb_profiles_bpic} compares $\mathit{SB}$ profiles
for the $\beta$~Pic systems in several gas models.
It shows that the nominal amount of gas leads to almost the
same profile as the one without gas (close to $-3.5$...$-4.0$). 
Furthermore, and surprisingly, all profiles with high and very high gas masses
(``ths'', ``tvs'', and ``thf'') are all close to each other and
to those without or with little gas
($\approx -3.5$...$-4.0$).

Our ``tvs'' and ``000'' models are close, although not completely identical, to
the ``high gas'' and ``no gas'' models in \citet{thebault-augereau-2005}.
In those two cases they obtained the slopes of $\approx -2.5$ 
and $\approx -4.0$, respectively (their Figs.~3 and~4).
Thus our result ($\approx -3.0$ and $\approx -3.5$...$-4.0$)
slightly differs from theirs.
Most of the difference comes from the fact that they used a more
realistic, extended distribution of parent planetesimals, but a much simpler,
monosized collisional model without a collisional disruption threshold.

Next, we compared the ``rns'' profile with the ``tns'' one.
Both turned out to be very similar (Fig.~\ref{fig_sb_profiles_bpic_p}).
This implies that both thermally-supported
and radiation pressure-supported gas disks with the same amount of gas
may yield similar radial distributions of dust.

Finally, we have also tested the influence of the radial profile of the gas temperature
on the $\mathit{SB}$ profiles. To this end, we compared the $\mathit{SB}$ profiles of $\beta$~Pic
obtained in the ``tns'' model ($p=1/2$)
with those in the ``t0ns'' and ``t1ns'' models ($p=0$ and $1$, respectively).
Similarly, ``rns'' results were compared with ``r0ns'' and ``r1ns''.
Again, the $\mathit{SB}$ profiles turned out to be almost indistinguishable (Fig.~\ref{fig_sb_profiles_bpic_p}).

\subsection{Surface brightness profiles with gas: all systems}

\begin{figure*}[bth!]
  \centerline{\includegraphics[width=0.75\textwidth]{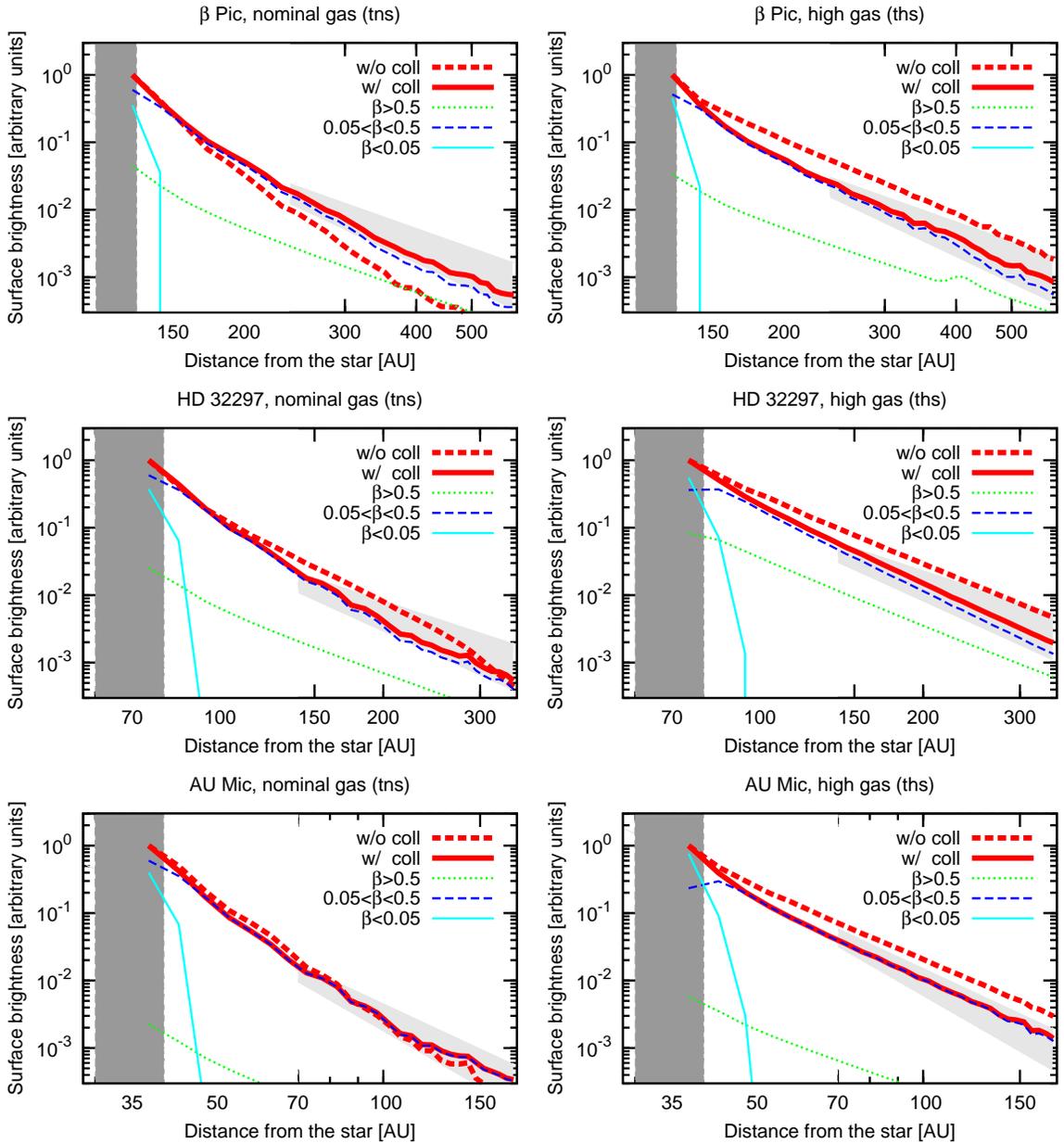}}
  \caption
  {
  $\mathit{SB}$ profiles for $\beta$~Pic (top), HD~32297 (middle), and AU~Mic (bottom),
  for a nominal gas model (left) and a high gas one (right).
  Thick dashed and solid lines are without and with collisions taken into account, respectively.
  Their normalization is the same as in Fig.~\ref{fig_sb_profiles_nogas}.
  Thin curves show partial contribution of three $\beta$-intervals into the total profile
  with collisions.
  Note that shaded areas have another meaning than in 
  Figs.~\ref{fig_sb_profiles_nogas}--~\ref{fig_sb_profiles_bpic_p}.
  They now indicate an approximate range of profiles, as deduced from observations (see Sect. 2):
  $-3.0$...$-4.0$ for $\beta$~Pic,
  $-2.7$...$-3.7$ for HD~32297,
  and $-3.8$...$-4.7$ for AU~Mic.
    \label{fig_sb_profiles}
  }
\end{figure*}

We now proceed with the numerical runs for all three systems
and two gas models (``tns'' and ``ths'') for each system (Table~\ref{tab_systems}).
The resulting profiles are shown in Fig.~\ref{fig_sb_profiles}.
One lesson from the plots is about the role of collisions.
In systems with a high dust density (or a large optical depth) and low gas
density, collisions flatten the profile ($\beta$~Pic, nominal gas; see also
Fig~\ref{fig_sb_profiles_nogas} without gas).
When the dust density is lower (AU~Mic), collisions have little influence on
the $\mathit{SB}$ slope. The same is true in the case of a high gas density
(high gas models for all three systems). This is 
mostly because the strong gas drag sustains sufficiently high radial
velocities even far from the star, so that $v_r(r)$ does not decrease
with increasing $r$ as abruptly as in the nominal gas cases
(see Fig.~\ref{fig_rad_vel}).

In addition, Fig.~\ref{fig_sb_profiles} depicts
partial contributions to $\mathit{SB}$ profiles by different-sized grains.
The largest contribution typically comes from medium-sized
grains with $0.05 < \beta < 0.5$, most of which
have stability distances outside the disk (Fig.~\ref{fig_r_stab})
and can be treated as ``effectively unbound''ones.
The relative contribution of the small $\beta > 0.5$ grains slightly
rises with increasing distance, but never becomes comparable to
that of the medium-sized particles.
Large grains with $\beta < 0.05$ do not make
any appreciable contribution to the $\mathit{SB}$ profiles in any of the systems.

However, the most important conclusion from Fig.~\ref{fig_sb_profiles}
is that the slopes differ only moderately for the two extreme gas models,
an effect that we have already seen for $\beta$~Pic and now see
for the other two systems.
In the nominal gas case, a distinctive feature of the profiles
is their slight ``curvature''~---  they do not follow
a single power-law across the entire disk.
This effect mostly comes from collisions rather than gas, and it can also
be seen in the gasless case (Fig.~\ref{fig_sb_profiles_nogas}).
The profiles are steeper close to the birth ring and are more gently
sloping farther out.
Between 2 and $3 r_\mathrm{birth}$ ($r_\mathrm{birth}$ being the birth ring distance),
the slopes are $-4.6$ for $\beta$~Pic,
$-4.5$ for HD~32297,
and $-4.7$ for AU~Mic.
Outside $3 r_\mathrm{birth}$,
the slopes flatten to
$-3.5$,
$-3.2$,
and $-3.3$,
respectively.
In contrast, in the high gas case the curvature effect is only present
close to the birth ring. 
Outside $2 r_\mathrm{birth}$,
all three $\mathit{SB}$ profiles have slopes
in a $-3.6$...$-4.0$ range.

Finally, the model slopes have to be compared with
with the observed slopes:
$-3.0$...$-4.0$ ($\beta$~Pic),
$-2.7$...$-3.7$ (HD~32297),
and $-3.8$...$-4.7$ (AU~Mic).
From all these values and from Fig.~\ref{fig_sb_profiles},
it is hardly possible to judge whether nominal-gas or high-gas models
match observations better.
In particular, this depends on the radial zone of the disk considered.
Besides, one should keep in mind that our models rest on many simplifying assumptions
(as, for instance, grey isotropic scattering) and have limited
accuracy (e.g., contain some numerical noise).
Equally, the slopes retrieved from observations inherit
uncertainties from the data and are sensitive to the specific procedure
of data reduction 
(see, e.g., a discussion in Sect. 3.2 of
\citeauthor{fitzgerald-et-al-2007b}
\citeyear{fitzgerald-et-al-2007b})
and should be treated with caution, too.
Thus the only conclusion we can make is that
nominal-gas and high-gas models are both is reasonable
agreement with observations.

%-----------------------------------------------------------------------
\section{Analytic model}

To better understand the numerical simulation results, in this section
we address the dust distributions analytically.

\subsection{``Static'' model in the case of a thermally-supported gas disk}

We start with a simple model that assumes dust to swiftly drift by gas
drag: $T_{\mathrm{drift}} \ll T_{\mathrm{life}}$, where
$T_{\mathrm{drift}}$ is the dust radial drift time
and $T_{\mathrm{life}}$ is the grain lifetime, e.g. collisional one.
With this assumption, a dust grain of radius $s$ is expected to
``instantaneously'' arrive at an equilibrium distance (\ref{stability})
from the star $r$ (for brevity the subscript ``stab'' will be omitted).
This model ignores that fact that grains spend finite time on their way
to stability distances, and thus also contribute to the optical depth
and brightness closer to the star than their parking distances.
It also ignores that stability distances of smaller grains
(if $p<1$) are located outside the disk and they never arrive there.
And, even if stability distances
are well inside the disk, grains may not arrive there (or not all of them
may), if they are collisionally eliminated on shorter timescales, meaning
that the assumption $T_{\mathrm{drift}} \ll T_{\mathrm{life}}$ fails.

Some of these assumptions fail in the systems considered in this paper.
For instance, in Sect. 5.5 we showed that the largest contribution to
the $\mathit{SB}$ profiles comes from ``effectively unbound'' grains.
Nevertheless, we deem this ``static'' model useful.
First, the ``static'' case is fully tractable analytically.
Second, it can give us a rough idea, to what extent the
ambient gas can change the dust profiles and it can be considered as
a limiting case for a more realistic ``dynamic'' model that will be worked out later.

The grains with radii $[s, s+ds]$ are produced in the birth
ring at a constant rate $\dot{N} ds$, see Eq.~(\ref{Ndot}).
These grains drift to a size-dependent stability distance, which is
easy to find.
In the limit of geometric optics, $\beta \propto s^{-1}$, Eq.~(\ref{beta}).
Next, $\eta$ is given by Eq.~(\ref{eta}).
Equating $\beta$ and $\eta$ we find a simple relation between the grain 
radius and stability distance (one-to-one if $p \ne 1$):
\be
  s \propto r^{p-1}.
\label{s_of_r}
\ee
Grains with radii $[s, s+ds]$ are located in an annulus $[r, r+dr]$.
Their steady-state number is
\be
  dN = \dot{N} T_{\mathrm{life}} ds,
\ee
and the normal geometrical optical depth of the annulus is
\be
  \tau (r)
  = \left| dN \over 2 \pi r dr \right| \pi s^2
  = {1 \over 2} {s(r)^2 \over r} \dot{N}(s(r)) T_{\mathrm{life}} (s(r)) \left| ds \over dr \right| .
\label{tau_general}
\ee
We now assume that the grain lifetime is independent of size (or distance).
This is a reasonable approximation as can be seen, for instance,
from Fig.~\ref{fig_T_coll}.
Substituting Eqs.~(\ref{s_of_r}) and (\ref{Ndot}) into Eq.~(\ref{tau_general}) yields
a radial dependence
\be
  \tau (r)
  \propto
  T_{\mathrm{life}} \; r^{-\alpha}
\label{tau_general1}
\ee
where
\be
  \alpha = 5 - 3p + (p-1)q .
  \label{alpha_wo_coll}
\ee
According to Eq.~(\ref{mu_alpha}), the $\mathit{SB}$ slope is 
\be
  \mu = 7 - 3p + (p-1)q .
  \label{mu_wo_coll}
\ee

In the above derivations we adopted the gas temperature which is independent
of the dust distribution. However, as discussed above (see Eq.~\ref{gamma}),
in a high dust density limit the gas temperature may become proportional
to a certain power of the dust density
(or equivalently the optical depth $\tau$ divided by distance $r$).
Using Eq.~(\ref{gamma}) instead of Eq.~(\ref{T_g}),
Eq.~(\ref{alpha_wo_coll}) for the optical depth slope replaces by
\be
  \alpha = {5 - 3p + (p-1)q - (3-q)\gamma \over 1 + \gamma(3-q)} ,
 \label{alpha_wo_coll_mod}
\ee
and Eq.~(\ref{mu_wo_coll}) for the $\mathit{SB}$ slope by
\be
  \mu = 2 + {5 - 3p + (p-1)q - (3-q)\gamma \over 1 + \gamma(3-q)}
 \label{mu_wo_coll_mod}
\ee

\begin{figure}[bth!]
  \vspace*{5mm}
  \centerline{\includegraphics[width=0.43\textwidth]{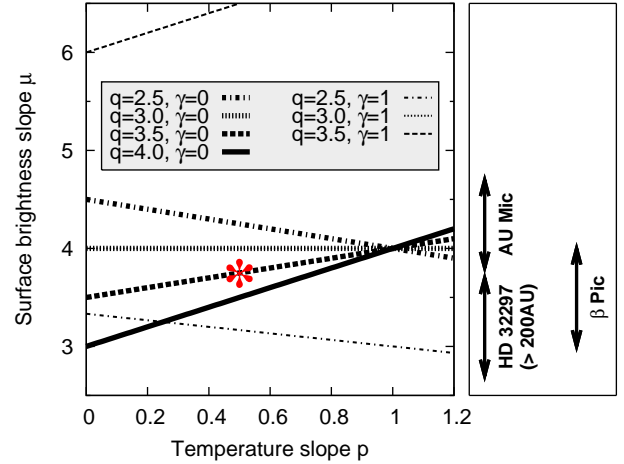}}
  \caption{
   Radial slope of surface brightness for different $p$, $q$, $\gamma$,
   according to Eq.~(\ref{mu_wo_coll_mod}).
   Thin and thick dotted lines ($q=3.0$) coincide with each other.
   Asterisk marks a standard model with
   a blackbody gas temperature ($\gamma=0$ and $p=1/2$) and
   a canonical size distribution of dust at production ($q=3.5$).
   For comparison, best-fit slopes inferred from the analysis of scattered light images
   of our three systems are shown on the right.
  \label{fig_sb_slope}
  }
\end{figure}

Figure~\ref{fig_sb_slope} gives slopes $\mu$ of surface brightness
(without grain-grain collisions)
for various $p$, $q$, and for $\gamma=0$ and $1$.
These results can easily be understood.
Consider, for instance, the standard gas heating model ($\gamma = 0$).
We start with the dependence on $q$ for a fixed $p$.
If the temperature gradient is not too steep ($p<1$),
then $s$ decreases with increasing $r$.
In that case, assuming a steeper size distribution (larger $q$)
makes the profile flatter, as it should.
For a steep temperature drop-off ($p>1$), it is the other way round.
The steeper the size distribution, the steeper the profile.
The dependence on $p$ for a fixed $q$ is also obvious.
For all $q > 3.0$ (which is expected), a steeper temperature gradient
steepens the $\mathit{SB}$ profile.
For all $q < 3.0$, the opposite is true.
In the standard case that we took in most of the numerical simulations
($p = 1/2$, $q = 3.5$),
we get $\mu = 3.75$ (asterisk in Fig.~\ref{fig_sb_slope}).

In the case of a dust-induced gas heating ($\gamma = 1$),
Eq.~(\ref{mu_wo_coll_mod})and Fig.~\ref{fig_sb_slope} suggest that
a much wider range of $\mu$ is possible, and that the results are
more sensitive to $q$. For example, the same standard case $p = 1/2$, $q = 3.5$
would result in an extremely steep $\mu = 6.5$.
However, in view of the temperature calculation results presented
in Fig.~\ref{fig_T_gas}, we do not expect Eq.~(\ref{mu_wo_coll_mod})
to be better approximation to reality than Eq.~(\ref{mu_wo_coll}).
Rather, it is meant to show that the radial distribution of dust
is quite sensitive to the assumed gas heating model.

\subsection{``Dynamic'' model}

Instead of considering a system in which all grains reside at their
stability distances, we now allow grains to drift through the disk
towards their respective parking positions.
We start with a thermally-supported gas disk and,
for a time being, we neglect collisions.
Equation~(\ref{tau_general})
for the normal geometrical optical depth of the disk
replaces by
\be
 \tau (r)
 =
 {1 \over 2r}
 \int_0^{s_0(r)} {\dot{N}(s) s^2 \over v_r(r,s)} ds 
 \propto
 {1 \over r}
 \int_0^{s_0(r)} {s^{2-q} \over v_r(r,s)} ds ,
\label{tau_unbound}
\ee
and the surface brightness profile is now given by
\be
  \mathit{SB} (r)
  \propto 
 {1 \over r^3}
  \int_0^{s_0(r)} {s^{2-q} \over v_r(r,s)} ds .
\label{mu_full}
\ee
Here, $s_0(r)$ is the radius of those grains whose stability distance is $r$.
For disks with $p<1$, $s_0$ decreases with $r$.
The radius $s_0$ or equivalently, the radiation pressure-to-gravity ratio of grains 
with radius $s_0$, which we denote by $\beta_0$,
can be read out from Fig.~\ref{fig_r_stab} or calculated analytically
from Eq.~(\ref{eta}):
\be
 \beta_0 (r)
 =
 (p+\xi)
 {k T_\mathrm{g}^0 \over \mu_\mathrm{g} m_\mathrm{H}}
 {r_0 \over GM_\star}
 \left( r \over r_0 \right)^{1-p} .
\label{boundary}
\ee
As a numerical example, for $\beta$~ Pic at $r=500$\,AU we have
$\beta_0 \approx 0.10$...$0.13$ and $s_0 \approx 6.6$...$8.6\mum$
for all thermally-supported disks modeled. 
It is only grains with $s \la 6.6$...$8.6\mum$ that are present
at the distance of $500$\,AU.

Unfortunately, $v_r(r,s)$ is a complicated, non-power-law function of both arguments
\citep[see, e.g., Eq. 23 in][]{takeuchi-artymowicz-2001}.
What is more,  a distance-dependent integration limit $s_0(r)$ in Eq.~(\ref{mu_full})
would result is a non-power-law brightness profile even if $v_r$ was a power law.
The only straightforward particular case is blowout grains with $\beta \ga 0.5$.
Fig.~\ref{fig_rad_vel} shows that their radial velocity is nearly constant,
yielding $\mathit{SB} \propto r^{-3}$. This is consistent with numerical simulations
(see dotted lines in Fig.~\ref{fig_sb_profiles}).

For the bound grains with moderate $\beta$ that actually dominate outer disks,
there are two competing effects.
One is the distance-dependent integration limit $s_0 (r)$.
Its physical meaning is as follows. The larger the distance, the smaller the grains
need to be to be able to reach it by gas drag. Thus larger grains are only
present closer to the star, which steepens the $\mathit{SB}$ slope.
Another effect is that $v_r$ decreases with $r$
(see Fig.~\ref{fig_rad_vel}), making the $\mathit{SB}$ slope flatter than $-3$.
For instance, if we adopt $v_r \propto r^{-1}$
as a very rough approximation for the outer disk, 
we will have $\mathit{SB} \propto r^{-2}$.

Further complications are expected from collisions. The grains have then
a limited lifetime, and the integrand in Eq.~(\ref{mu_full}) would have to be weighted
with a fraction of particles that survive collisions before they arrive at a distance $r$.
This would generally affect the $\mathit{SB}$ slope.

Because of the complexity of $v_r(r,s)$ and limited grain lifetime due to collisions,
it is difficult to extend our ``dynamic'' model further.
However, the model is still useful, as it uncovers the reason why $\mathit{SB}$ slopes in systems with gas
may be flatter than $-3$ \citep[like in][see their Fig.~3]{thebault-augereau-2005}:
it is the slow-down of radial drift velocity with increasing distance.
On the other hand, the model demonstrates that slopes steeper than $-3$ are also possible,
because larger grains can only drift to limited distances and only contribute to
the parts of the disk close to the birth ring.

Equations (\ref{tau_unbound})--(\ref{mu_full})
also hold  for the radiation pressure-supported disks. However, in this case
the upper integration limit $s_0$ has another meaning: the radius of
grains whose $\beta$ ratio is equal to $\beta_{\mathrm{gas}}$, Eq.~(\ref{eta rad}).
The above discussion applies in large part to this case, too.
The main conclusion is that a slope around $-3.0$ is expected,
possible deviations from which stem from a size-dependent radial
drift velocity and collisions.

%-----------------------------------------------------------------------

\section{Conclusions}

We considered young debris disks, in which there is observational evidence
for a rotating gas component (either primordial or secondary).
We assumed that dust is replenished from parent bodies that are located in a
``birth ring'' (which usually shows up in the resolved images).
We then modeled the dust distribution and scattered-light brightness profile
in the outer part of the disk, exterior to the birth ring, under different
assumptions about possible amount and distribution of gas.

Our main conclusions are as follows:

1. Our numerical simulations revealed that the radial profile
of dust density, and thus the surface brightness profile of a
dusty disk, are surprisingly insensitive
to the parameters of a central star,
location of the dust-producing planetesimal belt,
dustiness of the disk and,
most interestingly, the parameters of the ambient gas.
The radial brightness slopes in the outer disks are all typically
in the range $-3$...$-4$.
This result holds for gas densities varying by three orders of magnitude
and for different radial profiles of the gas temperature.
Both the gas of solar composition supported against gravity 
by gas pressure gradient
and gas of strongly non-solar composition that must be supported
by radiation pressure would lead to similar profiles, too.
The slopes of $-3$...$-4$ are the same
that were theoretically found for gas-free debris disks,
and they are the same as actually retrieved from observations
of many debris disks.

2. Although the slopes roughly fall into the range $-3$...$-4$,
the numerical simulations made it apparent that the exact slope
depends on the total amount of gas in the disk and the gas density
distribution slope, as well as on the dust density
(through collisional timescales).

3. We developed a simple analytic description of the radial
distribution of dust brightness in an optically thin disk.
The analytic model explains the numerical results 1.--2. and
provides guidelines to what can be expected in young debris disks,
prior to any detailed numerical modeling.
Assuming gray isotropic scattering by dust, the analytic model predicts
a range of slopes around $-3$, due to the dominant contribution
of small high-$\beta$ grains. It shows that deviations from this nominal value
may come from the slow-down of radial drift of bigger grains at larger distances (flattening),
from the fact that larger grains cannot reach larger distances (steepening),
and from the collisional elimination of dust particles.
In the limiting case of very high gas densities and low dust densities
(where the ``visible'' dust drifts through the outer disk
over timescales shorter than their collisional timescales),
if the dust size distribution at production follows
a power law with an index $-3.5$,
and assuming a black-body gas temperature,
the model predicts a slope of $-3.75$.

4. Our results for three young (10--30~Myr old), spatially resolved,
edge-on debris disks  ($\beta$~Pic, HD~32297, and AU Mic) show
that the observed radial profiles of the surface brightness do not pose
any stringent constraints on the gas component of the disk.
At least for gas densities falling within the
observationally derived density limits, we do not see any significant
effect of gas on dust distributions.
Thus we cannot exclude that
outer parts of the systems may have
retained substantial amounts of primordial gas
which is not evident in the gas observations
(e.g. as much as $50$ Earth masses for $\beta$~Pic).
However, the possibility that gas is only present in small to moderate
amounts, as deduced from gas detections
(e.g. $\sim 0.05$ Earth masses in the $\beta$~Pic disk or even less),
remains open, too.
In that case, gas would be secondary, stemming
for instance from grain-grain collisions or photo-desorption of dust.

%-----------------------------------------------------------------------
% Acknowledgments
%-----------------------------------------------------------------------

\acknowledgements
We thank Torsten L\"ohne for many useful discussions.
Comments by an anonymous referee that
helped to improve the paper are appreciated.
Part of this work was supported by the
\emph{Deut\-sche For\-schungs\-ge\-mein\-schaft, DFG\/} project number
Kr~2164/8-1,
by the {\em Deutscher Akademischer Austauschdienst}
(DAAD), project D/0707543, and by the International Space Science Institute
Bern, Switzerland (``Exozodiacal Dust Disks and Darwin'' working group,
http://www.issibern.ch/teams/exodust/).
FH was partly funded by the graduate student fellowship of the Thuringia State.
AB was funded by the \textit{Swedish 
National Space Board} (contract 84/08:1).

%-----------------------------------------------------------------------
% Appendix
%-----------------------------------------------------------------------

\appendix
\section{Nomenclature}

  \begin{tabular}{ll}
  Symbol	& Description \\
  \\
  \hline
  \\
  $d$			& Distance of the star from Earth\\
  $G$			& Universal gravitational constant\\
  $k$			& Boltzmann constant\\
  $n$			& Dust number density\\
  $q$			& Slope of dust size distribution in the birth ring (\ref{Ndot})\\
  $Q_D^\star$   	& Critical energy for fragmentation and dispersal (\ref{disruption})\\
  $Q_{\mathrm{pr}}$ 	& Radiation pressure efficiency \\
  $r$			& Distance from the star\\
  $L_\star$		& Stellar luminosity\\
  $L_\odot$		& Solar luminosity\\
  $m_\mathrm{H}$	& Mass of hydrogen atom\\
  $m_\mathrm{p}$	& Projectile mass (\ref{disruption})\\
  $m_\mathrm{t}$	& Target mass (\ref{disruption})\\
  $M_\star$		& Stellar mass\\
  $M_\odot$		& Solar mass\\
  $M_\oplus$		& Earth mass\\
  $s$			& Radius of a dust grain\\
  $\mathit{SB}$		& Surface brightness\\
  SED			& Spectral energy distribution\\
  $p$			& Radial slope of $T_\mathrm{g}$\\
  $P$			& Gas pressure\\
  $T_\mathrm{g}$	& Gas temperature\\
  $T_{\mathrm{coll}}$	& Collisional lifetime (\ref{T_coll})\\
  $T_{\mathrm{drift}}$	& Time a grain needs to reach its $r_{stab}$\\
  $T_{\mathrm{sett}}$	& Dust settling time (towards the disk mid-plane)\\
  $T_{\mathrm{stop}}$	& Dust stopping time (\ref{T_stop})\\
  $v_{\mathrm{cr}}$	& Minimum relative velocity for collisional disruption (\ref{disruption})\\
  $v_\mathrm{g}$	& Orbital velocity of a gas parcel\\
  $v_\mathrm{K}$	& Circular Keplerian velocity\\
  $v_r$			& Radial velocity of a dust grain\\
  $v_\mathrm{rel}$	& Relative velocity of two dust grains at collision\\
  $v_\mathrm{T}$	& Thermal velocity of gas\\
  $\alpha$		& Radial slope of $\tau$\\
  $\beta$		& Ratio of radiation pressure force to gravity for dust grains\\
  $\beta_{\mathrm{gas}}$& Ratio of radiation pressure force to gravity for gas\\
  $\gamma$		& Efficiency of gas heating by dust (\ref{gamma})\\
  $\eta$		& Ratio of gas-supporting force to gravity (\ref{eta def}), (\ref{eta rad})\\
  $\rho_\mathrm{bulk}$	& Bulk density of dust grains\\
  $\rho_\mathrm{g}$	& Gas density\\
  $\sigma$		& Collisional cross section of dust grains (\ref{T_coll})\\
  $\tau$		& Normal geometrical optical depth\\
  $\mu$			& Radial slope of $\mathit{SB}$\\
  $\mu_\mathrm{g}$	& Molecular weight of gas\\
  $\xi$			& Radial slope of gas density (\ref{xi})\\
  \\
  \hline
  \end{tabular}

%-----------------------------------------------------------------------
% References
%-----------------------------------------------------------------------

\newcommand{\AAp}      {A{\&}A}
\newcommand{\AApSS}    {AApSS}
\newcommand{\AApT}     {Astron. Astrophys. Trans.}
\newcommand{\AdvSR}    {Adv. Space Res.}
\newcommand{\AJ}       {Astron. J.}
\newcommand{\AN}       {Astron. Nachr.}
\newcommand{\ApJ}      {Astrophys. J.}
\newcommand{\ApJL}     {Astrophys. J. Lett.}
\newcommand{\ApJS}     {Astrophys. J. Suppl.}
\newcommand{\ApSS}     {Astrophys. Space Sci.}
\newcommand{\ARAA}     {Ann. Rev. Astron. Astrophys.}
\newcommand{\ARevEPS}  {Ann. Rev. Earth Planet. Sci.}
\newcommand{\BAAS}     {BAAS}
\newcommand{\CelMech}  {Celest. Mech. Dynam. Astron.}
\newcommand{\EMP}      {Earth, Moon and Planets}
\newcommand{\EPS}      {Earth, Planets and Space}
\newcommand{\GRL}      {Geophys. Res. Lett.}
\newcommand{\JGR}      {J. Geophys. Res.}
\newcommand{\MNRAS}    {MNRAS}
\newcommand{\PASJ}     {PASJ}
\newcommand{\PASP}     {PASP}
\newcommand{\PSS}      {Planet. Space Sci.}
\newcommand{\SolPhys}  {Sol. Phys.}
\newcommand{\SolSysRes}{Sol. Sys. Res.}
\newcommand{\SSR}      {Space Sci. Rev.}

%-----------------------------------------------------------------------

\end{document}